\documentclass[doublecol]{epl2} 

\usepackage{amsmath,amssymb,amsfonts} 
\usepackage{color}

\newcommand{\gwi}{gravitational-wave interferometer}
\newcommand{\Eqref}[1]{Eq.~\eqref{#1}}

\title{Interferometry of light propagation in pulsed 
fields}
\shorttitle{Interferometry of light propagation \ldots} 

\author{B. D\"obrich \and  H. Gies}
\shortauthor{B. D\"obrich \etal}

\institute{                    
Theoretisch-Physikalisches Institut, Friedrich-Schiller-Universit\"at Jena,\\
Max-Wien-Platz 1, D-07743 Jena, Germany
}


\abstract{
  We investigate the use of ground-based \gwi s for studies of the
  strong-field domain of QED. Interferometric measurements of phase velocity
  shifts induced by quantum fluctuations in magnetic fields can become a
  sensitive probe for nonlinear self-interactions among macroscopic
  electromagnetic fields. We identify pulsed magnets as a suitable
  strong-field source, since their pulse frequency can be matched perfectly
  with the domain of highest sensitivity of \gwi s. If these interferometers
  reach their future sensitivity goals, not only strong-field QED phenomena
  can be discovered but also further parameter space of hypothetical
  hidden-sector particles will be accessible.}

\begin{document}

\maketitle

\section{Introduction}

Charged quantum fluctuations as predicted by quantum electrodynamics (QED)
induce nonlinear self-interactions of electromagnetic fields
\cite{Heisenberg:1936qt}. This fundamental violation of the superposition
principle of classical electrodynamics has not yet been observed on the level
of macroscopic electromagnetic fields. Even though light-by-light interactions
have been verified in experiments involving high-energy photons
\cite{Akhmadaliev:1998zz}, an investigation of nonlinear interactions of
macroscopic fields would probe QED and its vacuum structure in a
large-amplitude regime which is comparatively little explored in quantum field
theory. In fact, large-amplitude or strong-field experiments not only give
access to unprecedented fundamental tests of QED, but also facilitate a search
for hypothetical particles with light masses and weak couplings to photons
(hidden-sector searches). This prospect has recently triggered a remarkable
growth of experimental and theoretical activities concerned with strong-field
and optical set-ups, for recent reviews see \cite{Dunne:2008kc,Gies:2008wv}.

A sensitive probe for vacuum nonlinearities is light propagation in strong
electromagnetic fields. The lowest-order nonlinear modifications of Maxwell's
theory as induced by QED vacuum polarization are described by the
(lowest-order) Heisenberg-Euler Lagrangian \cite{Heisenberg:1936qt},
\begin{equation}
  \mathcal{L} = \frac{1}{2}(\mathbf{E}^{2}-\mathbf{B}^{2})
  +\frac{2\alpha^{2}}{45m^{4}}(\mathbf{E}^{2}-\mathbf{B}^{2})^{2}
  +7\frac{2\alpha^{2}}{45m^{4}}(\mathbf{E}\cdot \mathbf{B})^{2}\ ,\label{eq:EHL}
\end{equation}
where the scale of nonlinearities is set by the electron mass $m$,
$\alpha\simeq 1/137$ denotes the fine-structure constant, and we use
$\hbar=c=1$. The resulting field equations predict that a plane wave in a
magnetic field $\mathbf B$ in vacuum propagates at a reduced phase and group
velocity \cite{Baier,Dittrich:2000zu},
\begin{equation}
  v_{i}=1-\frac{a_{i}}{45}\frac{\alpha^{2}}{m^{4}}B^{2}\sin^{2}\theta\
  ,\quad i=\bot,\parallel, \quad a_{\parallel}=14,\,\, a_{\bot}=8,
\label{eq:v_EH}
\end{equation} 
where $\theta$ denotes the angle between the $\mathbf B$ field and the
propagation direction. There are two propagation eigenmodes polarized parallel
$\|$ or perpendicular $\bot$ to the plane spanned by $\mathbf B$ and the
propagation direction. As the numerical coefficients $a_{\|,\bot}$ differ for
the two polarization modes, the magnetized quantum vacuum is birefringent.  A
number of experiments have already been carried out
\cite{Cameron:1993mr,Zavattini:2005tm,Bregant:2008yb,Chen:2006cd,battesti} or designed
\cite{OSQAR,Heinzl:2006xc} to look for vacuum birefringence in terms of
high-sensitivity polarimetry. The sensitivity limits achieved so far are
roughly four orders of magnitude above those necessary for the QED effect.

An alternative to polarimetry is given by absolute phase velocity measurements
by interferometry using, e.g., \gwi s, as first suggested by \cite{Boer:2002zw}, see
also \cite{Denisov:2004ep}. As \gwi s reach their highest sensitivity for
optical-path variations at a frequency $f\sim 10^2$Hz, the requirements for
large-scale magnet systems are enormous, as has recently been discussed in
detail in a concrete proposal in \cite{Zavattini:2008cr}. 

Indeed, using \gwi s for the search for nonlinear vacuum phenomena involves
many parameters which need to be taken into account in an optimized
fashion. In this work, we propose the use of pulsed magnets, as they are
developed and used in a stable manner in a number of laboratories worldwide at
field strengths approaching 100 Tesla. We derive the strain of the optical
path induced by repeated magnet pulses, as this is the central observable at
\gwi s. We also compute the signal-to-noise ratio for selected interferometers
as a criterion for measurability.  As we will demonstrate in the following,
not only strong-field QED phenomena can be discovered in such set-ups but also
further parts of the parameter space of hypothetical hidden-sector particles
will be accessible
.

\section{QED-induced strain in \gwi s}

Gravitational wave interferometers consist of two perpendicular interferometer
arms of equal length $L$. Mirrors at the arms' ends form an evacuated cavity
for two in-phase laser beams one for each cavity.  An incoming perturbation
such as a gravitational wave leads to a relative change $\Delta L$ of the
length of the interferometer arms, which manifests itself as a phase
difference of the two laser beams.  The {\em strain} $h(t)=\Delta L/L$
then corresponds to the amplitude of the incoming perturbation. Remarkably,
present day interferometers are, in certain frequency ranges, sensitive to a
strain of $h\sim10^{-22}$; future upgrades aim at further improvements of one
or two orders of magnitude.

In the context of quantum-induced refractive properties of the vacuum, it is
not the arm length but the corresponding optical path length $L$ which can be
modified by a strong field. If an external field is applied in a region $x<L$
in one interferometer arm, an optical-path difference $\Delta L$ is induced,
$(\Delta L/x) = 1-v$.  From now on, we concentrate on the parallel mode for
which the velocity shift is maximal. For this mode, the resulting strain is
\begin{eqnarray}
h(t) &=& \frac{\Delta
  L}{L}(t)
= \frac{x}{L}\, (1-v(t))
=\frac{x}{L}\frac{a_{\parallel}}{45}\frac{\alpha^{2}}{m^{4}}B(t)^{2} \\
&\approx & \frac{x}{L}(9.3 \times
10^{-24})\left(\frac{B(t)}{[1\text{T}]}\right)^{2}  \nonumber
.\label{eq:numerical_strain}
\end{eqnarray}
For a sizable strain, the magnet-length-to-detector-arm ratio $x/L$ should be
maximized. As $x$ is constrained by the technical realizability of the
magnetic field coil and the detector sensitivity requires long arm lengths, a
suitable compromise between detector arm length and sensitivity has to be
found.

\section{Detector sensitivity}

The detector sensitivity to the relative shift of the length of the
interferometer arms $\Delta L/L$ is limited by various sources of noise. At
low frequencies $f\lesssim40\mathrm{Hz}$, the main limitation arises from
seismic activities for ground-based detectors. For instance at $f\sim
  10$Hz, which seems ambitious but feasible for big dipole magnets, the
  sensitivity measure $S_{h}(f)$ (as detailed below) is suppressed by more than three
  orders of magnitude compared to the peak sensitivity and depletes rapidly
  for even smaller frequencies.
At intermediate and higher
frequencies, thermal and shot noise, respectively, limit the detector
sensitivity \cite{Saulson:1995zi,Blair:1991wd}
.
We conclude that observing the quantum-induced phenomena
requires a magnetic field modulation in the optimal frequency range for a
given detector. Pulsed magnets which can be modulated over an appropriate
parameter range therefore appear to be good candidates.

Information about the optimal frequency range is encoded in the total spectral
density function $S_{h}(f)$ of the noise, see e.g. \cite{Blair:1991wd}. For our
estimates, we concentrate on advanced LIGO (in preparation) and GEO600
(operational).
The projected sensitivity of both detectors slightly depends on the
  details of the event acting as a source for the interferometric signal. In
  the case of LIGO, we use the strain sensitivity for neutron-star binaries
  \cite{ligocurves} for our calculations, which is satisfactory over a wide
  range of intermediate frequencies (for a typical expected sensitivity curve
  of the advanced LIGO, see \cite{Smith:2009bx}). For GEO600, we use the
typical sensitivity data available at \cite{geocurves}, which can be well
approximated by a fit function of the form
\begin{equation}
  S_{h}(f) =  S_{0}((f_{0}/f)^{p_{1}}+2(f/f_{0})^{p_{2}}+2)/5
  ,
\end{equation} 
where $f_{0}=560\mathrm{Hz}, S_{0}=7\times10^{-44}\mathrm{Hz}^{-1}$,
$p_{1}=3.8$, $p_{2}=3$ near the sensitivity maximum. For the advanced LIGO, no simple fit is available.
As the signal induced by QED or other hypothetical particles can be predicted exactly, we expect
  that the interferometer sensitivity can even be optimized for the present
  phenomenon.

As a measure for the observability of a shift of the optical path $\Delta
L/L$, we determine the signal-to-noise ratio (SNR) $d$ of the induced
strain. In the present case, the SNR equals the expectation value of the
detector output divided by the standard deviation of the output variable due
to noise.
Using a matched filter (or ``Wiener filter'') for the signal, the SNR
$d$ for a gravitational wave interferometer is given by (see
e.g. \cite{Blair:1991wd,Schutz:1995hn,Thorne:1987af} and references therein)
\begin{equation}
d^{2}=2\int_{0}^{\infty}\frac{|\tilde{h}(f)|^{2}}{S_{h}(f)}\mathrm{d}f\
,\quad \tilde{h}(f)=\int_{-\infty}^{\infty}h(t)e^{-2\pi ift}\mathrm{d}t,
\label{eq:SNR_def}
\end{equation}
where $\tilde{h}(f)$ is the Fourier transform of the induced strain.

\section{Pulsed magnetic fields\label{sub:Pulsed}}

As an example, we consider pulsed fields that can be obtained at the
Dresden High Magnetic Field Laboratory (HLD)
\cite{wosnitza}. The HLD aims at providing $100\mathrm{T}$
fields generated by a solenoid in a non-destructive set-up, i.e., the
infrastructure is maintained and the experiment can in principle be repeated
arbitrarily often.

As the magnetic pressure is given by $p_{\mathrm{mag}}=B^{2}/2\mu_{0}$,
already at fields strengths of about $B=50\mathrm{T}$, the pressure on the
coils is four orders of magnitude above the atmospheric pressure, which
requires a careful coil design. As a consequence, the coils are usually
heavily mantled and it is difficult to render the interferometer laser beam
orthogonal to the external magnetic field within a single coil setup. In order
to maximize the shift of the optical path, a pair of Helmholtz coils has to be
used instead. In this manner the laser beam of the
interferometer can be aligned in parallel to the field coils and thus mainly
orthogonal to the magnetic field lines without interfering with the coil
mantle.

For such ``split coils'', a maximum field strength of up to $60\mathrm{T}$
appears technically feasible, at a coil diameter of about $x=0.2\mathrm{m}$ and
a coil separation of $\mathcal{O}(1\mathrm{cm})$. As the beam waist of the
interferometer lasers is of the order of $\mathrm{cm}$, the interferometer
beam can fit between the magnet coils, even though the issue of stray photons
may require further discussion which goes beyond the scope of the present
work.  Also for standard Helmholtz setups, the field is roughly constant at a
sizable extent only along the direction of the magnetic field lines, whereas
the quantum-vacuum effect requires a sizable field length perpendicular to the
field lines; the length of the latter is of the order of the coil
separation. For the proposed setup, the coil design thus needs to be optimized
to provide for high (but not necessarily constant) magnetic field strengths,
spatially extending orthogonally to the direction of the field lines.

A typical pulse undergoes a damped oscillation with pulse frequency
$\nu_{\mathrm{B}}$ and damping rate $\gamma$. For $N$
subsequent pulses at times $t_{0}\dots t_{N-1}$, a satisfactory
description is given by
\begin{equation}
  B(t)=B_{0}\sum_{i=0}^{N-1}\theta(t-t_{i})
  \sin(2 \pi \nu_{B}(t-t_{i}))\exp(-\gamma(t-t_{i})).
  \label{eq:model_pulse}
\end{equation} 
Here, we have ignored that successive pulses have no temporal overlap in a
single-magnet set-up. However, as the pulse repetition rate
$\nu_{\mathrm{P}}\equiv1/(t_{i+1}-t_{i})$ is much smaller than the damping
rate (see below), Eq. \eqref{eq:model_pulse} is a well justified
approximation.  The pulse frequency $\nu_{B}$ in Eq. \eqref{eq:model_pulse}
depends on the total capacity of the capacitor banks and can lie in the range
$\mathcal{O}(\mathrm{ms}\dots\mathrm{s})$, while the damping rate $\gamma$ is
mainly determined by the heat capacity of the coil.  In addition, the
achievable pulse repetition rate $\nu_{\mathrm{P}}$ in a non-destructive
mode depends strongly on the desired peak field strength.  An ambitious, but
nevertheless feasible Helmholtz setup should be able to achieve a maximum
field strength of $B_{\mathrm{max}}=60\mathrm{T}$, followed by a reverse field
of $B_{\mathrm{min}}=-6\mathrm{T}$ and thus a damping to about 10\% of the
peak field strength
\footnote{It can be expected that also pulses with damping to about 70\% will be obtainable in the near future. The above pulse parameters are in that sense conservative, since a smaller damping factor $\gamma$ leads to a higher strain (cf. \eqref{eq:h(f)square}}.
This choice fixes the
amplitude of the model pulse  (Eq. \eqref{eq:model_pulse})
$B_{0}\approx148\mathrm{T}$ and implies the constraint
\begin{equation}
\gamma=2 \nu_{B} \ln\left|\frac{B_{\mathrm{max}}}{B_{\mathrm{min}}}\right|
\label{eq:gamma_omega}\ .
\end{equation}
We use the remaining free parameter $\nu_{B}$ for optimizing the SNR
\eqref{eq:SNR_def} (within the technical limitations).  For this,
we need the modulus of the Fourier transform of the strain. Ignoring overlap
terms of successive pulses as argued above, we find
\begin{eqnarray}
&{}&|\tilde{h}(f)|^{2}=\frac{x^{2}}{L^{2}}(9.3 \times
10^{-24})^2\left(\frac{B_0}{[1\mathrm{T}]}\right)^4  \label{eq:h(f)square} \\
&{}&\times 
\frac{(\pi \nu_{B})^{4}
  \sin^{2}\left(\pi\frac{f}{\omega_{\mathrm{P}}}N\right)
  \csc^{2}\left(\pi\frac{f}{\omega_{\mathrm{P}}}\right)
}{(\gamma^{2}+\pi^{2}f^{2})(\gamma^{4}+\pi^{4}(f^{2}-4\nu_{B}^{2})^2
+2 \gamma^{2}\pi^{2}(f^{2}+4\nu_{B}^{2}))}  \nonumber
.
\end{eqnarray}
For a single pulse $N=1$, the trigonometric functions in
Eq. \eqref{eq:h(f)square} cancel and the dependence on the repetition rate
$\nu_{\mathrm{P}}$ drops out, as expected. For a large number of pulses
$N$, the trigonometric functions yield a representation of a $\delta$
comb, 
\begin{equation}
  \sin^{2}\left(\pi\frac{f}{\nu_{\mathrm{P}}}N\right)
  \csc^{2}\left(\pi\frac{f}{\nu_{\mathrm{P}}}\right)\approx
  N\sum_{n\in\mathbb{N}}\delta\left(\frac{f}{\nu_{\mathrm{P}}}-n\right).
\label{eq:sin_csc}
\end{equation}
At large $N$, only frequencies which are multiples of the pulse repetition
rate $\nu_{\mathrm{B}}$ thus contribute to the SNR \eqref{eq:SNR_def}. As
$\nu_{\mathrm{B}}$ is much smaller than the frequencies dominating the SNR,
the contributing frequencies form a quasi-continuum such that the sum in
Eq. \eqref{eq:sin_csc} can well be approximated by an integral. As a result,
the square of the SNR for $N$ pulses can be expressed in terms of the
single-pulse result to a good accuracy:
\begin{equation}
d^{2}|_{N}\approx N\ d^{2}|_{1}\ .\label{eq:SNRmult_SNR}
\end{equation}
The reproducibility of the signal by non-destructive pulsed magnets thus is a
lever arm for an enhancement of the SNR by a factor of $\sqrt{N}$.  For the advanced LIGO with $L=4000\mathrm{m}$, the
sensitivity curve has a broad minimum of the order of $S_h(f)\approx10^{-47}\mathrm{Hz}^{-1}$ for frequencies ranging from approximately $50$Hz to $500$Hz. Maximizing $d^{2}$ by varying
the pulse parameter $\nu_{B}$ yields $\nu_{B}\approx47\mathrm{Hz}$,
implying $\gamma\approx217\mathrm{Hz}$ by means of \Eqref{eq:gamma_omega}.
Inserting these values into Eq. \eqref{eq:SNR_def}, we obtain the SNR for a
single pulse,
\begin{equation}
d|_{1}^{\mathrm{LIGO}}\approx 1.9\times 10^{-2}.\label{eq:d_ligo}
\end{equation}
As a result, about $N\approx2763$ pulses are required in order to
achieve a total SNR of $\mathcal{O}(1)$. Depending on the details of the
setup, an SNR of $\mathcal{O}(10)$ might eventually be required. For the
following feasibility study, however, we only demand for an SNR of
$\mathcal{O}(1)$. This is also justified because the expected signal can be
predicted to a high accuracy which will allow for an adapted noise filtering.

As mentioned above, the re-cooling time for the magnet system which determines
the pulse-repetition rate depends mainly on the pulse energy.  A realistic
estimate lies in the order of several minutes, implying a continuous operation
of the facility for a few days. This appears perfectly feasible.  GEO 600 is
considerably less sensitive than advanced LIGO but, for our purposes, profits
from the shorter arm length of 600m. Maximizing $d^2$ with respect to
$\nu_{B}$ yields $\nu_{B}\approx273\mathrm{Hz}$ with
$\gamma\approx1259\mathrm{Hz}$ and thus a pulse length below
$1\mathrm{ms}$. As a result, $N\approx2.3\times10^{6}$ pulses are necessary to
observe the QED induced strain at GEO. This corresponds to an unrealistic
measurement time of a few years. In consequence, GEO in combination with
presently available pulsed magnets is not well suited for the observation of
the QED induced strain. Nevertheless, it still has a new-physics discovery
potential, see below.

\section{Search for hidden-sector particles}

The investigation of vacuum nonlinearities can also be used to search for
hypothetical particles with small masses and weak couplings to photons
\cite{Gies:2008wv}. An apparatus designed to observe the QED effect will also
explore a significant range of a new-physics parameter space. Here, we
concentrate on potential velocity shifts induced by axion-like particles
(ALPs) \cite{Maiani:1986md} or minicharged particles (MCPs)
\cite{Gies:2006ca}.

\begin{figure}
\begin{center}
\includegraphics[width=1\linewidth]{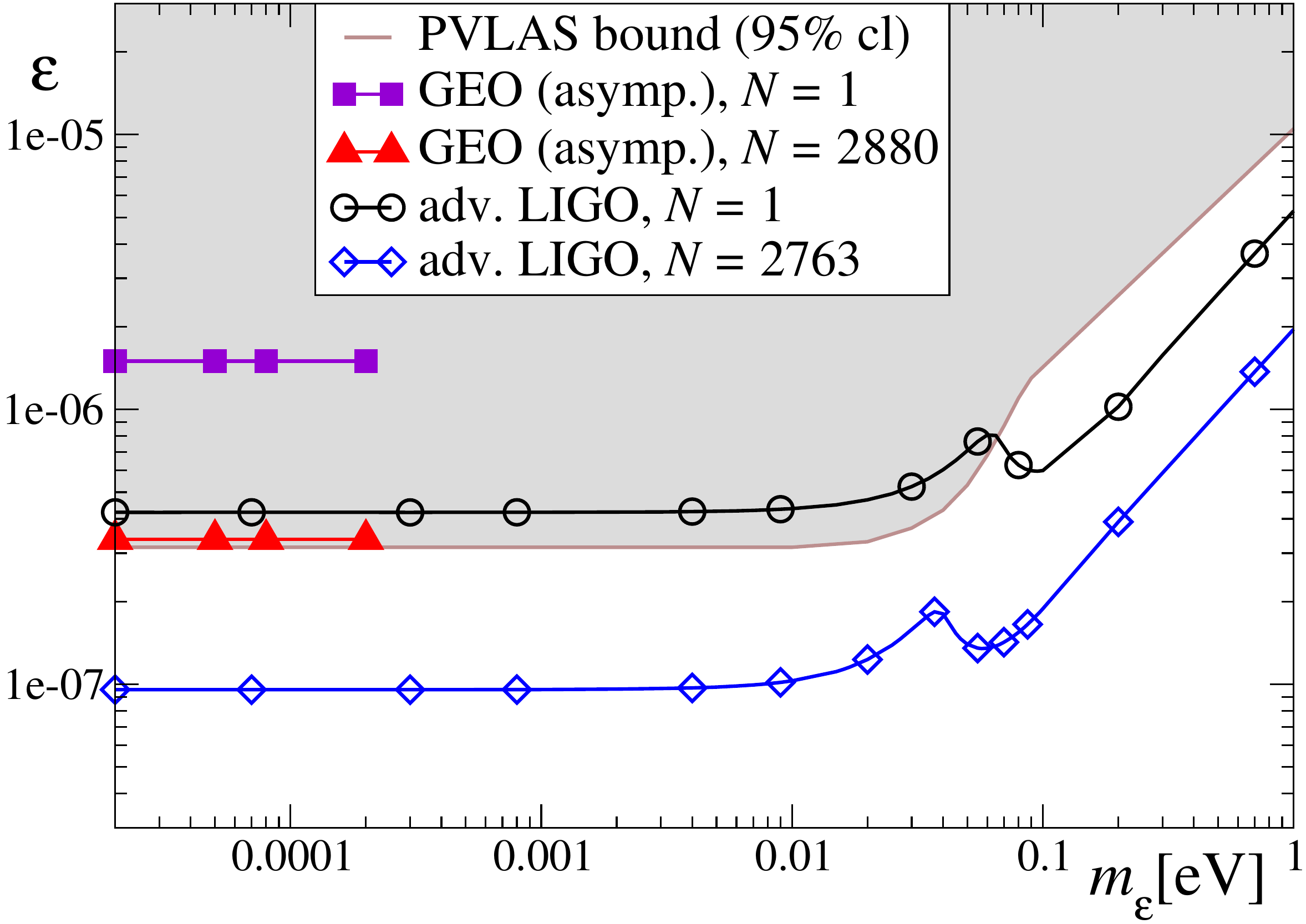}
\caption{New-physics discovery potential for spin-$\frac{1}{2}$ minicharged
  particles (MCP) in the coupling-mass plane. For MCPs, already a
  single-pulse ($N=1$) measurement at advanced LIGO (or $N\simeq2880$ at GEO)
  can approach or slightly improve the current best laboratory bounds from
  PVLAS. The use of $N=2763$ pulses as needed for the QED effect can lead to
  sizable improvements over the whole mass range.} 
\label{fig:MCP}
\end{center}
\end{figure}

MCPs with charge $Q=\varepsilon e$ and mass $m_{\varepsilon}$ induce vacuum
effects analogous to electrons. As the MCP mass can be much lighter than the
electron, quantum-induced velocity shifts have to be calculated to all orders
in the field strength parameter $\varepsilon e B/m_\varepsilon^2$ and the
frequency $\omega/m_\varepsilon$ \cite{Tsai:1975iz,Ahlers:2006iz}.  MCP masses
corresponding to a Compton wavelength larger than the volume inside the
Helmholtz coil cannot be fully resolved within the setup. Thus, the coil
separation of $\mathcal{O}(1\mathrm{cm})$ constrains the search for MCP masses
to $m_{\varepsilon}\gtrsim 2\times10^{-5}\mathrm{eV}$.  Using the formulas of
\cite{Gies:2006ca,Ahlers:2006iz} for a Dirac spinor MCP (scalar MCPs behave
similarly \cite{Ahlers:2006iz}), we calculate the effect for the polarization
component parallel to the external field, which maximizes the velocity shift
as in the QED case.  For an interferometer laser with $\omega=1.2\mathrm{eV}$
and the pulse shape as used for the QED effect, we obtain exclusion limits in
the fractional charge-mass plane $(\varepsilon,m_\varepsilon)$ by demanding a
SNR of $\mathcal{O}(1)$, see Fig.\ref{fig:MCP}. These
exclusion limits display two characteristic limits which correspond to the two
asymptotic limits of the velocity shift: $1-v\sim \varepsilon^4
B^2/m_\varepsilon^4$ for large masses (cf. \Eqref{eq:v_EH}) and $1-v\sim
-\varepsilon^{8/3} B^{2/3}/\omega^{4/3}$ for small masses.

\begin{figure}
\begin{center}
\includegraphics[width=1\linewidth]{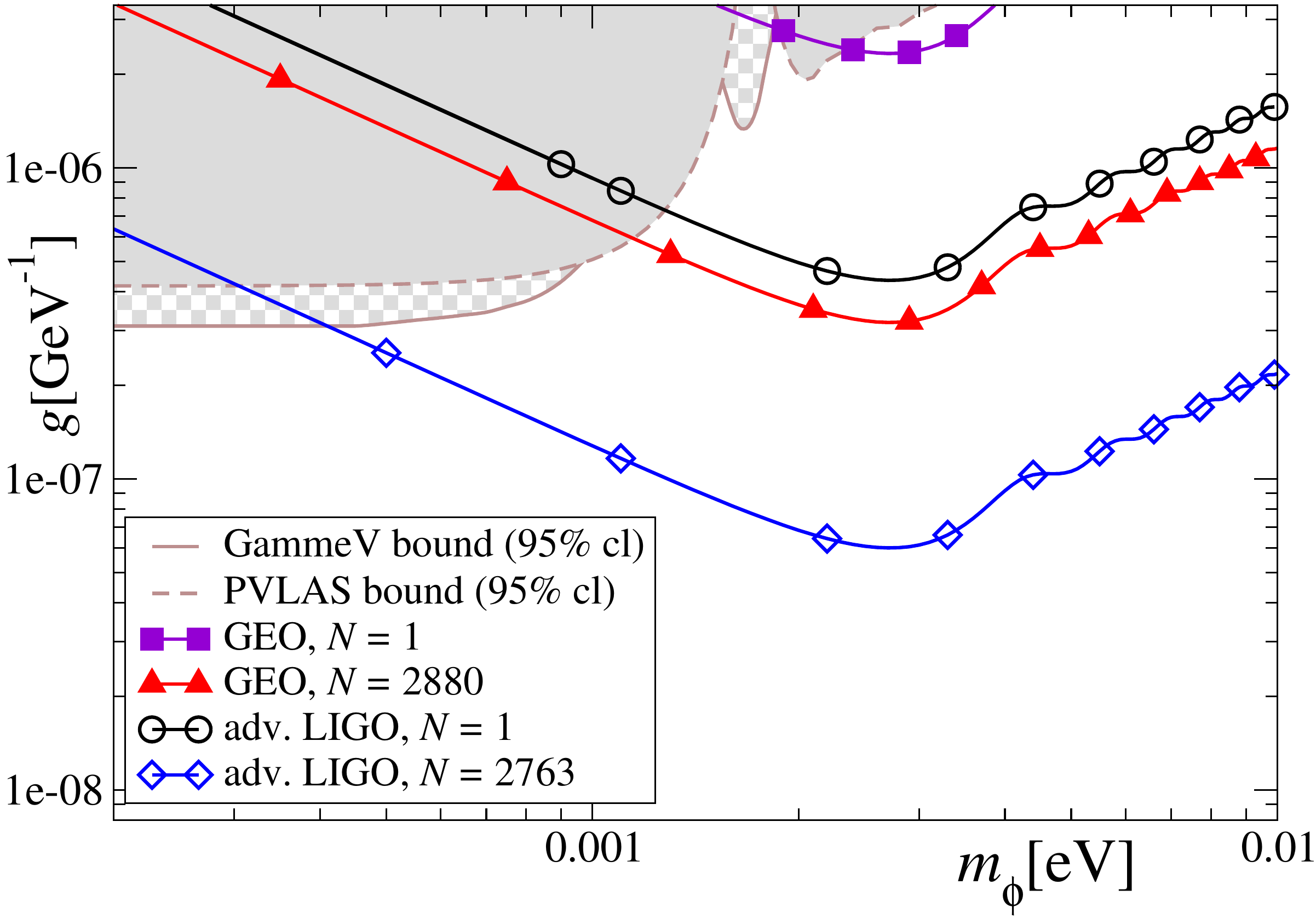} 
\caption{
  This figure shows the discovery potential in the coupling-mass plane for axion-like particles (ALP)
  at LIGO and GEO, respectively.
  For ALPs, the current best laboratory bounds by GammeV and PVLAS can be improved
  significantly in the meV mass range and above by advanced LIGO as well as
  already by GEO.}
\label{fig:ALP}
\end{center}
\end{figure}

We find that already a single-pulse ($N=1$) measurement at advanced LIGO can
approach the current best laboratory bounds \cite{Ahlers:2006iz} derived from
PVLAS data \cite{Zavattini:2005tm} with potential minor improvements in the
larger-mass range. Assuming a measurement time of ten days at GEO with a
magnet re-cooling time of 5 minutes, implying $N\simeq 2880$ pulses would
arrive at a similar bound. The small-mass asymptotics of the GEO bounds for
$N=1$ and $N=2880$ are also shown in Fig.~\ref{fig:MCP}.
Advanced LIGO with $N=2763$ as needed for QED can reach a sensitivity
of $\varepsilon\simeq 10^{-7}$ for $m_\varepsilon\lesssim0.01$eV. This
would compete with current cosmological bounds \cite{Melchiorri:2007sq}. Of
course, astrophysical energy loss considerations can lead to much stronger
bounds \cite{Davidson:2000hf} reaching down to $\varepsilon \sim
10^{-14}$. However, as stellar physics probes a much larger momentum-transfer
regime, these bounds depend on the underlying microscopic physics and thus may
not apply to laboratory experiments \cite{Jaeckel:2006xm}.
We conclude that an investigation of the QED effect can also improve current
laboratory bounds on MCPs to an unprecedented level.

Axion-like particles (ALPs) are uncharged scalars (S) or pseudo-scalars (P)
with a mass $m_{\phi}$ and an effective ALP-photon interaction of
the form
$\mathcal{L}_{\mathrm{P}}=g\phi_{\mathrm{P}}\mathbf{E\cdot B}$
or
$\mathcal{L}_{\mathrm{S}}=\frac{1}{2}g\phi_{\mathrm{S}} (\mathbf{B}^2 - \mathbf{E}^2)$
with coupling
$g$. The interaction structure implies that scalars only couple
to the $\perp$ mode, whereas pseudo-scalars couple to the $\|$ mode of the
propagating light in a magnetic field. Loosely speaking, an effective
reduction of the phase velocity arises from the fact that the corresponding
photon partly propagates as a massive ALP component. The corresponding
velocity shifts read \cite{Maiani:1986md}
\begin{equation}
1-v_\|^{\text{P}} =1-v_\bot^{\text{S}} 
=\frac{B^{2}}{2m_{\phi}^{2}}g^{2}
\left(1-\frac{\sin(2y)}{2y}\right)\ ,\quad 
y=\frac{xm_{\phi}^{2}}{4\omega},\label{eq:ior_alp}
\end{equation}
where $\omega$ denotes the laser frequency, and $x$ is the length of the
magnet interaction region.  As discussed in detail in \cite{Ahlers:2006iz},
the polarimetry of strong-field particle searches can distinguish between the
various particle scenarios, see also \cite{Zavattini:2008cr}. For the current
study, we only concentrate on the velocity shift irrespective of the
polarization dependence for simplicity.

Let us first consider the ALP parameter range for the coupling $g$ and mass
$m_\phi$ that can be probed by a single pulse.  Using the pulse shape as for
the QED case, we obtain the accessible region in the mass-coupling plane, see
Fig. \ref{fig:ALP} for a comparison with the current best
laboratory limits from PVLAS and GammeV \cite{Chou:2007zzc} (see also the BMV
experiment \cite{Robilliard:2007bq}). Already a single-pulse measurement at
advanced LIGO can improve on existing bounds for masses above 1meV. For
$m_\phi\gtrsim 4\times10^{-4}\mathrm{eV}$, higher number of pulses as needed
for the QED effect give access to a parameter space which is significantly
larger than current laboratory limits. Again, astrophysical considerations and
observations lead to stronger bounds by a few orders of magnitude in the
$g\sim 10^{-11}\mathrm{GeV}^{-1}$ regime \cite{Zioutas:2004hi}, but may not be
directly applicable to laboratory experiments due to a different
momentum-transfer regime \cite{Jaeckel:2006xm}.

\section{Conclusion}

The sensitivity goal of ground-based \gwi s appears well suited to use these
experiments for exploring the strong-field domain of QED. In order to generate
a strongly magnetized quantum vacuum in such an interferometric experiment, we
have identified pulsed magnets as an advantageous strong-field source for two
reasons: they provide extremely strong laboratory magnetic fields, and their
pulse frequency can be perfectly matched with the region of highest
sensitivity of the \gwi s.

For our quantitative estimates, we have concentrated on the advanced LIGO
detector, as its sensitivity goal matches with currently available field
strengths already in a rather conservative estimate. Pushing the various
components to their limits may facilitate a detection also at the \gwi s,
which are currently operational such as GEO 600. Also the fact, that the
quantum-induced signal can be theoretically predicted to a good accuracy may
give rise to an improved noise filtering.

From a general perspective, the QED velocity shift as well as the MCP signal
in the large-mass domain and the dispersive ALP effect scale with $x B^2$,
where $B$ is the amplitude and $x$ the extent of the magnetic field. For the
use of \gwi s, also a suitable time variation of the magnetic field is
needed. Whereas pulsed fields profit from extremely high fields and a suitable
time variation, their deficit is a smaller extent in comparison to dipole
magnets. Since pulsed fields win roughly an order of magnitude in the field
strength and lose an order of magnitude in the field extent, the quantity
$xB^2$ can generically still be an order of magnitude larger for pulsed fields
than for dipoles. A similar consideration has inspired the development and use
of pulsed magnets in the BMV experiment \cite{battesti,Robilliard:2007bq}
which finally aims at a parameter goal of $xB^2\simeq
600\mathrm{T}^2\mathrm{m}$ (recent experimental results of BMV have been
achieved with $xB^2\simeq 40\mathrm{T}^2\mathrm{m}$) \cite{battesti}. The
pulsed Helmholtz coil configuration considered in this work which is inspired
by ongoing experiments at the Dresden High Magnetic Field Laboratory would
yield $xB^2\simeq 720\mathrm{T}^2\mathrm{m}$, which agrees with the design
goal of BMV. But whereas the time dependence of the field pulse at BMV can be
a disturbing factor for the BMV polarimetry, the time structure of the field
pulse is a necessary and advantageous key feature for a strong-field
experiment at a \gwi.

Given the prospect of exploring a new parameter regime of strong-field quantum
field theory with implications for the search for new elementary particles,
establishing a strong-field quantum-vacuum program at \gwi s appears to be
worthwhile.

\acknowledgments We would like to thank Achamveedu Gopakumar, Thomas
Herrmanns\-d\"orfer, J\"org J\"ackel, Guiseppe Ruoso, Gerhard Sch\"afer and
David Shoemaker for helpful discussion.  This work is
supported by DFG SFB-TR18 and under contract Gi 328/5-1 (Heisenberg program).

\end{document}